\documentclass[sigconf, authorversion]{acmart}

\AtBeginDocument{%
  \providecommand\BibTeX{{%
    \normalfont B\kern-0.5em{\scshape i\kern-0.25em b}\kern-0.8em\TeX}}}

\copyrightyear{2023} 
\acmYear{2023} 
\setcopyright{rightsretained} 
\acmConference[SC-W 2023]{Workshops of The International Conference on High Performance Computing, Network, Storage, and Analysis}{November 12--17, 2023}{Denver, CO, USA}
\acmBooktitle{Workshops of The International Conference on High Performance Computing, Network, Storage, and Analysis (SC-W 2023), November 12--17, 2023, Denver, CO, USA}
\acmDOI{10.1145/3624062.3624214}
\acmISBN{979-8-4007-0785-8/23/11}

\usepackage{listings}
\usepackage{xcolor}
\usepackage{multicol}
\usepackage{subcaption}
\newcommand{\ket}[1]{|#1\rangle}

\begin{document}

\title{TISCC:  A Surface Code Compiler and Resource Estimator for Trapped-Ion Processors}

\author{Tyler LeBlond}
\affiliation{%
  \institution{Computational Sciences and Engineering Division \\ Oak Ridge National Laboratory}
  \city{Oak Ridge}
  \country{USA}}
\email{leblondtr@ornl.gov}

\author{Justin G. Lietz}
\affiliation{%
  \institution{National Center for Computational Sciences \\ Oak Ridge National Laboratory}
  \city{Oak Ridge}
  \country{USA}}
\email{lietzjg@ornl.gov}

\author{Christopher M. Seck}
\affiliation{%
  \institution{Computational Sciences and Engineering Division \\ Oak Ridge National Laboratory}
  \city{Oak Ridge}
  \country{USA}}
\email{seckcm@ornl.gov}

\author{Ryan S. Bennink}
\affiliation{%
  \institution{Computational Sciences and Engineering Division \\ Oak Ridge National Laboratory}
  \city{Oak Ridge}
  \country{USA}}
\email{benninkrs@ornl.gov}

\begin{abstract}
    We introduce the Trapped-Ion Surface Code Compiler (TISCC), a software tool that generates circuits for a universal set of surface code patch operations in terms of a native trapped-ion gate set. To accomplish this, TISCC manages an internal representation of a trapped-ion system where a repeating pattern of trapping zones and junctions is arranged in an arbitrarily large rectangular grid. Surface code operations are compiled by instantiating surface code patches on the grid and using methods to generate transversal operations over data qubits, rounds of error correction over stabilizer plaquettes, and/or lattice surgery operations between neighboring patches. Beyond the implementation of a basic surface code instruction set, TISCC contains corner movement functionality and a patch translation that is implemented using ion movement alone. Except in the latter case, all TISCC functionality is extensible to alternative grid-like hardware architectures. TISCC output has been verified using the Oak Ridge Quasi-Clifford Simulator (ORQCS).
\end{abstract}

\maketitle

\section{Introduction}
\label{sec:introduction}

The surface code is the most popular quantum error-correcting code (QECC) due to its high error threshold ($\sim 1\%$) and nearest-neighbor connectivity, which improves its applicability to real hardware~\cite{fowler2012surface}. It is therefore important for the broader community to focus on developing surface code compilation frameworks, and several authors have proposed such frameworks (either abstractly or in software) using so-called lattice surgery, which is the leading technique for implementing fault-tolerant operations with surface codes~\cite{horsman2012surface,fowler2018low,litinski2019game,beverland2022surface,beverland2022assessing,paler2020opensurgery,watkins2023high}. While compilers that transform quantum circuits into fault-tolerant operations on the surface code exist~\cite{paler2020opensurgery,watkins2023high}, we are not aware of any compilers that transform the latter into explicit hardware circuits. Such a tool is not only essential to an end-to-end fault-tolerant compilation pipeline for quantum circuits, which might be used once quantum computing hardware reaches the requisite level of maturity, but it is also helpful in the near-term for the purposes of:

\begin{enumerate}
    \item resource estimation for fault-tolerant implementations of quantum algorithms using a realistic hardware model,
    \item co-design of the fault-tolerant and hardware layers of the quantum computing stack,
    \item simulation of logical error rates for fault-tolerant operations using error channels derived from a realistic hardware model,
    \item and developing explicit workflows for translating measurement outcomes into values of logical operators.
\end{enumerate}

To this end, we introduce the Trapped-Ion Surface Code Compiler (TISCC), which is a software tool that generates hardware-level circuits and resource estimates for surface code patch operations on trapped-ion processors~\footnote{The open-source repository can be found at https://github.com/ORNL-QCI/TISCC.}. By combining TISCC with other tools that compile logical circuits to surface code instructions, an end-to-end pipeline for compilation and resource analysis can be realized. TISCC utilizes a realistic trapped-ion hardware model; in compiling surface code operations into hardware instructions, it explicitly accounts for the movement of ions between trapping zones on a two-dimensional grid of repeating units which themselves consist of sites serving as either memory sites, operation sites, or junctions. Given a compiled hardware circuit, TISCC ensures its validity by simulating ion movements on the grid and resolving junction conflicts. The TISCC hardware model also counts space-time resources required for surface code operations by tracking the nominal time at which each hardware operation should occur in the final circuit using literature-derived estimates of the duration of each hardware instruction and accounting for parallelism. Members of a universal instruction set for surface code quantum computing are built out of a set of verified surface code patch primitive operations, including transversal operations over data qubits, rounds of error correction over stabilizer plaquettes, lattice surgery operations between pairs of patches, corner movements, and an operation that swaps columns of data qubits left-ward.

As was argued in Ref.~\cite{LeBlond2023ASCR}, the development of open-source fault-tolerant compilers that are both extensible and verified is timely. Recent demonstrations of fault-tolerance using the surface code on real hardware (e.g. Ref.~\cite{google2023suppressing}) make surface code compilation software a present need. In this vein, many different surface code compilation methodologies have been proposed, but little work has been done in benchmarking them using software implementations. Therefore, in the near term, the extensibility of compilers to different visions of surface code compilation could enable the exploration of resource trade-offs between these visions. As mentioned above, TISCC uses a set of verified primitives to implement members of a higher-level surface code instruction set. These combinations of operations are verified by extension as long as the primitives are correctly used. Although we have not done so here, the verified primitives available in the TISCC library can aid in the implementation of alternative instruction sets, making TISCC extensible to other surface code compilation methodologies and useful for benchmarking them.

Additionally worthy of note is that recent progress in quantum low-density parity check (QLDPC) codes with higher encoding rates and and better code distance scaling than the surface code begs the question of its longevity as the preferred QECC~\cite{breuckmann2021quantum}. That said, roadblocks to implementing QLDPC codes involve the fact that their Tanner graphs often include non-local connections that make them difficult to embed in a 2D grid-like hardware architecture, though recent work has explored the decomposition of their Tanner graphs into planar subgraphs, potentially improving their implementation feasibility~\cite{bravyi2023high,tremblay2022constant}. While TISCC has been written specifically for surface codes, it has been written in a modular fashion with extensibility to different grid-like hardware architectures and logical qubit implementations on those architectures in mind. We foresee that the functionality of TISCC could be expanded to QLDPC codes as they become practical.

This paper is organized as follows: In Sec.~\ref{sec:surface_codes}, we introduce the surface code lattice surgery model targeted by TISCC in a way that is abstracted from the trapped-ion hardware model. In Sec.~\ref{sec:hardware_model}, we describe the trapped-ion hardware model employed by TISCC, explain how surface codes are mapped onto hardware in this model, and describe our approach to resource estimation using TISCC-generated circuits. In Sec.~\ref{sec:verification}, we describe our method for verifying TISCC circuits using quasi-Clifford simulation and process tomography in the logical sub-space. Finally, in Sec.~\ref{sec:conclusion}, we conclude by providing an outlook for surface code compilation and the future of TISCC. In the appendices, we include alternative surface code instructions that were not included in the main text (Sec.~\ref{sec:derived}) and some details about the software design and implementation (Sec.~\ref{sec:software_design}).

\section{Abstract Surface Code Lattice Surgery Model}
\label{sec:surface_codes}
In this section, we will introduce the fault-tolerant instruction set that we have in mind when we discuss surface code compilation. We will also discuss some details of operation for these instructions themselves in order to motivate a particular set of surface code primitives that we implement within TISCC. This discussion will be abstracted from the details of our chosen hardware model, which will be discussed in Sec.~\ref{sec:hardware_model}. We will assume that the reader knows some basics about quantum computing and the surface code.

\subsection{Motivation for the Choice of a Local Tile-Based Lattice Surgery Instruction Set}
Firstly, one must appreciate that there are (at least) two steps to surface code compilation. In the first step, logical circuits are compiled to a surface code instruction set, and in the second step, surface code instructions are compiled to hardware instruction set. This means that there are at least three layers to the stack, each having both its own instruction set and entities (abstract or real) on which its instruction set acts. The top layer, which is the most familiar, consists of logical gates acting on logical qubits. In the middle (fault-tolerant) layer, logical qubits are encoded in large blocks of physical qubits as part of a QECC. If purposed to serve as more than a quantum memory, this QECC ought to have a known set of fault-tolerant operations that forms a complete instruction set for universal quantum computation using their encoded logical information. The bottom layer of the stack is the hardware layer, the discussion of which we reserve for Sec.~\ref{sec:hardware_model}.

For surface codes (our chosen QECC), there have been many propositions for instruction sets, but there appears to be consensus that the lowest-overhead instruction set uses so-called lattice surgery on surface code patches, contiguous regions of hardware that each encode one logical qubit~\cite{horsman2012surface, fowler2018low}\footnote{Recent authors using twist-based lattice surgery have occasionally supported the use of patches that encode two logical qubits each~\cite{litinski2019game, beverland2022assessing}. We avoid twist-based lattice surgery in this work (see Sec.~\ref{sec:twists} for an explanation).}. In lattice surgery, edge-wise merge and split operations between adjacent patches are used to produce entanglement between logical qubits. This enables the compilation of CNOT gates, which in turn enables the compilation of T and S gates using typical state teleportation circuits~\cite{fowler2012surface} and the ability to inject appropriate resource states~\cite{li2015magic, lao2022magic, gavriel2023transversal}. Together with the Hadamard gate, which can be accomplished through a transversal Hadamard followed by patch rotation (if necessary)~\cite{vuillot2019code,beverland2022surface,litinski2019game, fowler2018low}, these gates complete the Clifford+T gate set, which is universal for quantum computing.

\label{sec:local_lattice_surgery}
\begin{table*}[htbp]
  \centering
  \caption{Local lattice surgery instruction set implemented using TISCC primitives. All instructions in this set are implied to act on (and return) one or two logical tiles as defined in Sec.~\ref{sec:logical_tile}.}
  \begin{tabular}{|l|p{9cm}|c|c|}
    \hline
    \textbf{Instruction} & \textbf{Description} & \textbf{Logical Tiles In/Out} & \textbf{Logical Time-Steps} \\
    \hline
    Prepare X/Z & Initializes one uninitialized tile to a $\ket{+}$ or $\ket{0}$ state fault-tolerantly & 1 & 1 (0) \\
    Inject Y/T & Initializes one uninitialized tile to a $\ket{Y}$ or $\ket{T}$ state non-fault-tolerantly &  1 & 0 \\
    Measure X/Z & Measures one initialized tile in the X/Z basis and makes uninitialized & 1 & 0 \\
    Pauli X/Y/Z & Applies logical Pauli operator to an initialized tile & 1 & 0 \\
    Hadamard & Performs a transversal Hadamard gate over an initialized tile & 1 & 0 \\ 
    Idle & Performs $d_t$ cycles of error correction on an initialized tile & 1 & 1 \\
    Measure XX/ZZ & Measures the joint XX/ZZ operators of two vertically/horizontally-adjacent initialized tiles & 2 & 1 \\ 
    \hline
  \end{tabular}
  \label{tab:surf_code_instr_set}
\end{table*}

Explicitly compiling and estimating resources for a limited yet universal set of surface code operations was a guiding motivation for the development of TISCC. However, some popular lattice surgery approaches rely on Pauli product measurements with support on an arbitrary number of logical qubits~\cite{litinski2019game, beverland2022assessing}, which leads to an unbounded set of possibilities. This approach, often called sequential Pauli-based computation~\cite{beverland2022surface}, also results in a serialization of logical circuits that we wished to avoid in our conceptual framework. For these reasons, we decided that a \textit{local} lattice surgery instruction set that was compatible with a direct Clifford+T compilation strategy would be desirable. The surface code instruction set that has been targeted by TISCC includes instructions acting on one or two neighboring units of area on an extended two-dimensional grid. We call these units of area \textit{logical tiles} in order to abstract them from their hardware implementation and to consider them as a resource analogous to the number of fault-tolerant clock cycles required to execute a quantum circuit, commonly referred to as \textit{logical time-steps} in surface code compilation references. Many other references on surface code compilation have considered the notion of tiles in the fault-tolerant layer; e.g. Refs.~\cite{litinski2019game,beverland2022surface, beverland2022assessing, watkins2023high, chamberland2022universal}. However, as far as we know, others have not considered tiles---as opposed to patches or logical qubits---to be the fundamental entities within the fault-tolerant layer. 

A significant motivation for using a tile-based instruction set is that tiles are a natural basis for the placement\footnote{As we are concerned with tile-to-hardware compilation rather than circuit-to-tile compilation, we do not address tile layout or macro architecture here; see \cite{litinski2019game} for an approach that minimizes spatial resources.} and scheduling of lattice surgery operations.
While quantum circuits nominally concern logical qubits only, data movement and/or non-local operations such as CNOT involve ancillary resources that do not necessarily correspond to qubits from the logical circuit and which must be tracked to avoid resource contention.
As a case in point, a protocol to implement CNOT operations between diagonally adjacent tiles, using a third tile between them, was presented in~\cite{horsman2012surface}.
Ref.~\cite{beverland2022surface} generalized this protocol to implement long-range CNOT operations via Bell states encoded in intermediary patches connecting the targeted qubit tiles.
We find that tiles are convenient units for mapping out patches, identifying paths between them, and tracking free and busy regions of the hardware.
Moreover, long-range operations between remote patches can be conveniently implemented in just two time steps using parallel local tile-based operations.
In one step, local tile-based operations create a chain of local Bell states along a path of tiles connecting the targets.  In a second step, a set of Bell measurements along the chain propagate entanglement to the chain ends, effectively performing the CNOT.

\subsection{Local Lattice Surgery Instruction Set}
\begin{figure}[htbp]
    \centering
    \includegraphics[width=0.85\columnwidth]{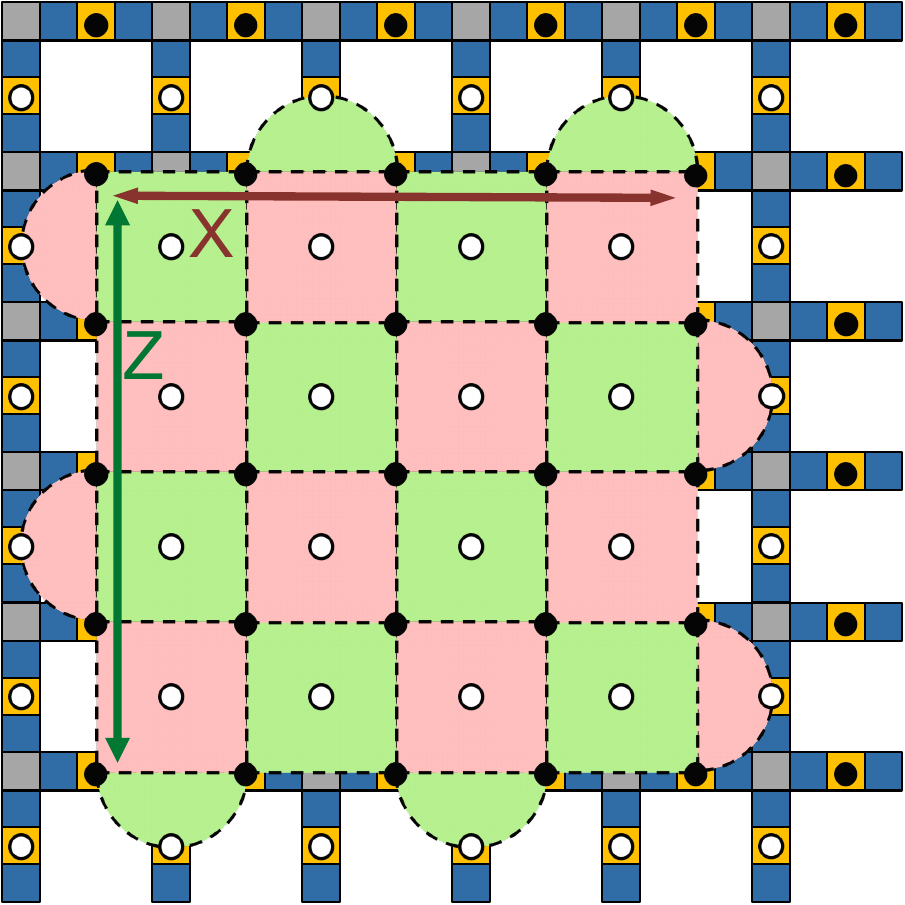}
    \captionsetup{skip=10pt}
    \caption{Pictorial representation of a surface code patch in what we call the `standard arrangement' of stabilizers, with X stabilizers colored red (dark) and Z stabilizers colored green (light). The patch is superimposed over a section of the trapped-ion hardware that we abstractly call a `logical tile', with memory (`M') sites colored dark blue, operations (`O') sites colored gold, and junction (`J') sites colored grey. Sites are shown occupied by data qubits (black circles) and measure qubits (white circles). }
    \label{fig:standard_arrangement}
\end{figure}

Having established that we intend to use a \textit{local}, \textit{tile-based} surface code lattice surgery instruction set, here we consider the necessary elements of this set and how these elements lend themselves to our choice of verified primitives. Our chosen set (see Table~\ref{tab:surf_code_instr_set}) is very similar to the one that was presented in Ref.~\cite{beverland2022surface}, though with a few modifications. 
We have demoted the local Bell preparation, Bell measurement, and Move instructions to what we call `derived' instructions (they can be built from instructions in Table~\ref{tab:surf_code_instr_set} but have more efficient implementations, see Sec.~\ref{sec:derived}). We include explicit logical Pauli X/Y/Z operators, though these are typically thought of as being tracked in software using the Pauli frame~\cite{fowler2012surface,riesebos2017pauli}, and state injection circuits since they are required in order to complete a universal instruction set. Our Hadamard instruction is transversal, and though we do not include a subsequent rotation, TISCC contains functionality that could be used to this end (see Sec.~\ref{sec:rotation}). Finally, we note that our Prepare X/Z instructions require a full logical time-step as stand-alone instructions; this is to preserve the requirement that all instructions output valid tiles. However, as is often noted in the literature, state preparation operations can be combined with subsequent lattice surgery operations to take effectively take 0 time-steps; several of the instructions in Table~\ref{tab:surf_code_derived_instr} exploit this feature.  Interestingly, Table~\ref{tab:surf_code_instr_set} does not include any entangling gates (e.g. CNOT); instead, entangling operations are realized via entangling measurements (Measure XX/ZZ).

\begin{figure*}[htbp]
    \centering
    \includegraphics[width=0.9\textwidth]{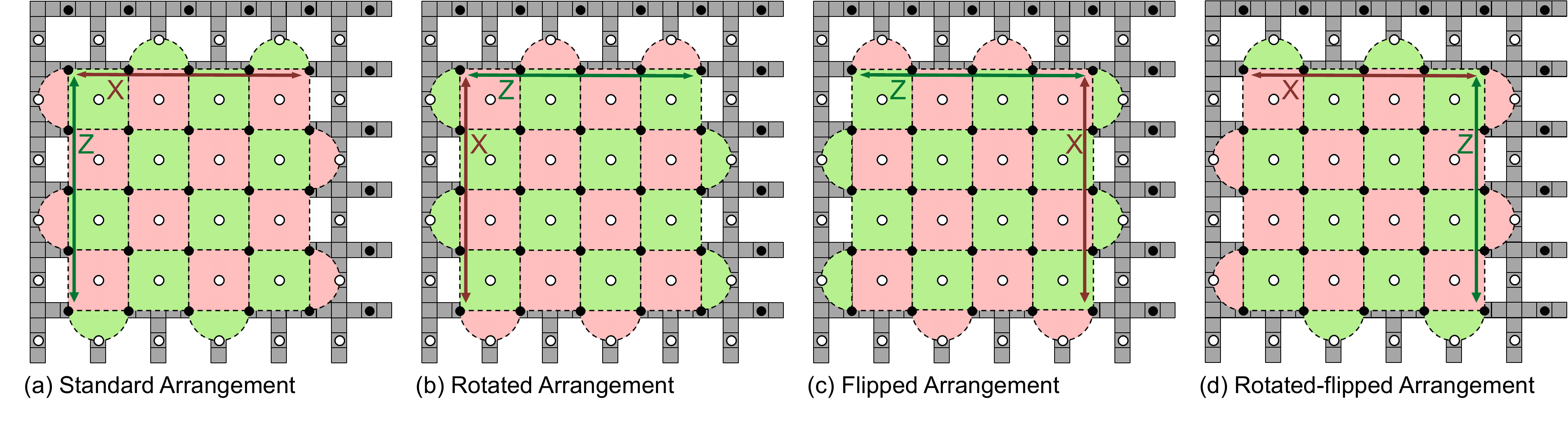}
    \captionsetup{skip=10pt}
    \caption{We refer to four canonical stabilizer arrangements (left to right): (a) the standard arrangement (same as Fig.~\ref{fig:standard_arrangement}), (b) the rotated arrangement, (c) the flipped arrangement, and (d) the rotated-flipped arrangement. All arrangements are accessible through combinations of the LogicalQubit::xz\_swap and LogicalQubit::flip\_patch member functions.}
    \label{fig:canonical_arrangements}
\end{figure*}

\subsection{Definition of Logical Tile}
\label{sec:logical_tile}

The logical tile is an abstraction of the hardware area capable of supporting a single surface code patch encoding one logical qubit with X and Z code distances $d_x$ and $d_z$. Knowing the elements of our local, tile-based lattice surgery instruction set, we now define our logical tile and the constraints that this definition imposes. Firstly, notice that each instruction in Table~\ref{tab:surf_code_instr_set} specifies either one or two logical tiles as input and one or two logical tiles as output, and that these logical tiles are specified as either initialized or uninitialized in the description. In our conception, the `initialized' or `uninitialized' status of a logical tile refers to whether or not there is an operable surface code patch occupying it. Our notion of operability is that a surface code patch consists of what we define as the `standard arrangement' of stabilizer plaquettes, which can be seen visually in Fig.~\ref{fig:standard_arrangement}. Typically, tiles become initialized through a Prepare X/Z operation or uninitialized through a Measure X/Z operation, but other derived instructions (see Table~\ref{tab:surf_code_derived_instr} in Sec.~\ref{sec:derived}) might also toggle their initialization status. While alternative stabilizer arrangements are available within TISCC (see Fig.~\ref{fig:canonical_arrangements} for what we call the four canonical arrangements), the elements of Table~\ref{tab:surf_code_instr_set} always leave surface code patches in the standard arrangement if an initialized tile is output\footnote{The Hadamard instruction, which leaves behind a rotated patch, is the exception to this rule.}. Additionally, while a merged patch (spanning two tiles) is an intermediate state in the Measure XX/ZZ instruction, these patches are always subsequently split. These considerations prevent the need to generalize our instruction set to act on patches spanning arbitrary numbers of tiles or on patches with arbitrary valid stabilizer arrangements, which aids in the reduction of the elements of Table~\ref{tab:surf_code_instr_set} into the verified primitives of Table~\ref{tab:surf_code_prim}.

There are further corollaries to our definition of logical tile that assist in the reduction of our instruction set into primitives. First, the standard arrangement implies that logical Z (X) operators run vertically (horizontally)\footnote{While all logical Z (X) operators are topologically equivalent to each other in the sense that they are the same up to products of stabilizers, in practice one must account for the signs of stabilizer measurement outcomes to map from one to another~\cite{fowler2018low}. In TISCC, there are notions of \textit{default}-edge and \textit{opposite}-edge logical operators for a patch, where the default-edge for the standard arrangement is depicted in Fig.~\ref{fig:standard_arrangement}. If new boundary stabilizers are measured, the qubits supported by a default-edge logical operator may change. TISCC includes functionality both to perform such transformations and track resulting operator deformations (see Sec.~\ref{sec:rotation})}. Therefore, vertical (horizontal) merges between patches in this arrangement correspond with XX (ZZ) measurements. In order to facilitate two-patch operations, one (two) strips of ancillary qubits are inserted between the patches to facilitate merge operations in the case of odd (even) code distances\footnote{TISCC includes functionality for even code distances although it is standard to only employ odd code distances for surface codes.}. While literature sources differ on whether an ancillary strip of physical qubits located in between surface code patches is necessary to perform lattice surgery operations (see Refs.~\cite{horsman2012surface, fowler2018low, de2020zx} and Refs.~\cite{litinski2019game, beverland2022assessing} for examples in either case), we make use of one in order to (a) avoid two boundary stabilizers ever occupying a unit of space that is typically serviced by one measurement qubit (see Sec.~\ref{sec:circuit_mappings} for details on how stabilizer circuits are executed on the hardware) and (b) enable logical XX and ZZ measurements to take place in only the time needed to perform the corresponding merge operation fault-tolerantly\footnote{It seems to be a common oversight in the literature that, without this ancillary strip, the post-split boundary stabilizers would require $d_t$ measurement cycles in order for them to be known fault-tolerantly, causing the full XX/ZZ measurement to take two surface code time steps instead of one.}. Thus, a logical tile in TISCC consists of $ 2 \lceil\frac{d_z+1}{2}\rceil$ rows and $ 2 \lceil\frac{d_x+1}{2}\rceil$ columns on the underlying hardware grid.

\subsection{Primitive Surface Code Operations}
\label{sec:prim_ops}
\begin{table*}[htbp]
  \centering
  \caption{Surface code primitive operations that TISCC uses to implement instructions from Table~\ref{tab:surf_code_instr_set} and Table~\ref{tab:surf_code_derived_instr}.}
  \begin{tabular}{|l|l|p{6.5cm}|c|c|}
    \hline
    \textbf{Name} & \textbf{Function Call} & \textbf{Description} & \textbf{No. Patches} & \textbf{Logical Time-Steps} \\
    \hline
    Prepare Z & LogicalQubit::transversal\_op & Initializes data qubits in the X/Z basis & 1 & 0 \\
    Measure Z & LogicalQubit::transversal\_op & Measures data qubits in the X/Z basis & 1 & 0 \\
    Hadamard & LogicalQubit::transversal\_op & Applies Hadamard gate to data qubits and swaps roles of X and Z operators & 1 & 0 \\
    Inject Y/T & LogicalQubit::inject\_state & Initializes data qubits according to desired state & 1 & 0 \\
    Pauli X/Y/Z & LogicalQubit::apply\_pauli & Applies Pauli gate according to logical operator & 1 & 0 \\
    Idle & LogicalQubit::idle & Performs $d_t$ cycles of error correction & 1 & 1 \\
    Merge & LogicalQubit::merge & Merges two patches into a larger patch in the standard arrangement & 2 & 1 \\ 
    Split & LogicalQubit::split & Splits a merged patch into two patches with their same pre-merge specification & 2 & 0 \\
    \hline
  \end{tabular}
  \label{tab:surf_code_prim}
\end{table*}

In Table~\ref{tab:surf_code_prim} we list the primitives used within TISCC to implement instructions from Table~\ref{tab:surf_code_instr_set} and Table~\ref{tab:surf_code_derived_instr}. Some members of Table~\ref{tab:surf_code_instr_set} are implemented using only one of these primitives, while others consist of combinations of them. In most cases TISCC contains functionality for applying these primitives to alternative stabilizer arrangements (see Fig.~\ref{fig:canonical_arrangements}) or deformed patches (see Sec.~\ref{sec:rotation}), but for the purpose of implementing Table~\ref{tab:surf_code_instr_set} and Table~\ref{tab:surf_code_derived_instr} they should be thought of as acting on patches in the standard arrangement of Fig.~\ref{fig:standard_arrangement}. The primitives in Table~\ref{tab:surf_code_prim} operate on patches rather than tiles, and do not need to output the same number of patches that they act on. They are not constrained to the specification of Sec.~\ref{sec:logical_tile}. Therefore, we emphasize that the set of primitives in Table~\ref{tab:surf_code_prim} is not itself our local, tile-based surface code instruction set. Instead, we use combinations of these primitives under the constraints of Sec.~\ref{sec:logical_tile} to implement our instruction set. It should also be noted that instructions other than those presented here can be implemented using the TISCC library, and that the verification of the primitives of Table~\ref{tab:surf_code_prim} is in most cases not limited to patches in the standard stabilizer arrangement (details on the verification process can be found in Sec.~\ref{sec:verification}). Having enumerated the primitives that were used to implement our instruction set, we now discuss further functionality available within TISCC. 

\subsection{Patch Deformations and Translations}
\label{sec:rotation}


In some cases, it may be desirable to perform a patch rotation immediately following a Hadamard instruction\footnote{A transversal Hadamard gate leaves the patch in a rotated stabilizer arrangement [Fig.~\ref{fig:canonical_arrangements}(b)], and the rotation returns the patch to the standard arrangement.}. There have been several different proposals for patch rotation protocols in the literature, each with a different required overhead~\cite{horsman2012surface, fowler2018low, vuillot2019code, litinski2019game, beverland2022surface}. While we have not yet implemented any of them here, we have included some of the enabling functionality for TISCC users to experiment with different trapped-ion implementations of patch rotations themselves.

To this end, methods are implemented within TISCC to deform patches by adding and removing boundary stabilizers. This results in the modification of logical operators existing along patch boundaries, which is typically called corner movement. Corner movements can be useful either within patch rotation protocols or to expose logical operators as desired for subsequent lattice surgery operations~\cite{litinski2019game, fowler2018low}. Crucial to the way that TISCC handles corner movements is the construction and maintenance of a parity check matrix by every LogicalQubit object. Using this matrix, a given boundary stabilizer is added by finding (and removing or replacing) the existing stabilizers and logical operators that anti-commute with it. Any logical operator with support on the added stabilizer is also updated in favor of its lower-weight counterpart. Modifications made to default-edge logical operators are tracked in order to aid users in reconstructing their values in classical post-processing (see Sec.~\ref{sec:post-processing}). Where necessary, TISCC also handles the measurement and/or preparation of corner qubits as needed to maintain a valid single-qubit patch. Corner movement functionality can be accessed through the LogicalQubit::extend\_logical\_operator\_clockwise method, which finds the sequence of boundary stabilizers that need to be measured in order to accomplish the desired movement. Currently, patch deformations within TISCC cannot change the number of logical qubits encoded by a patch and cannot add stabilizers other than boundary stabilizers. 

\begin{figure}[htbp]
    \centering
    \includegraphics[width=\columnwidth]{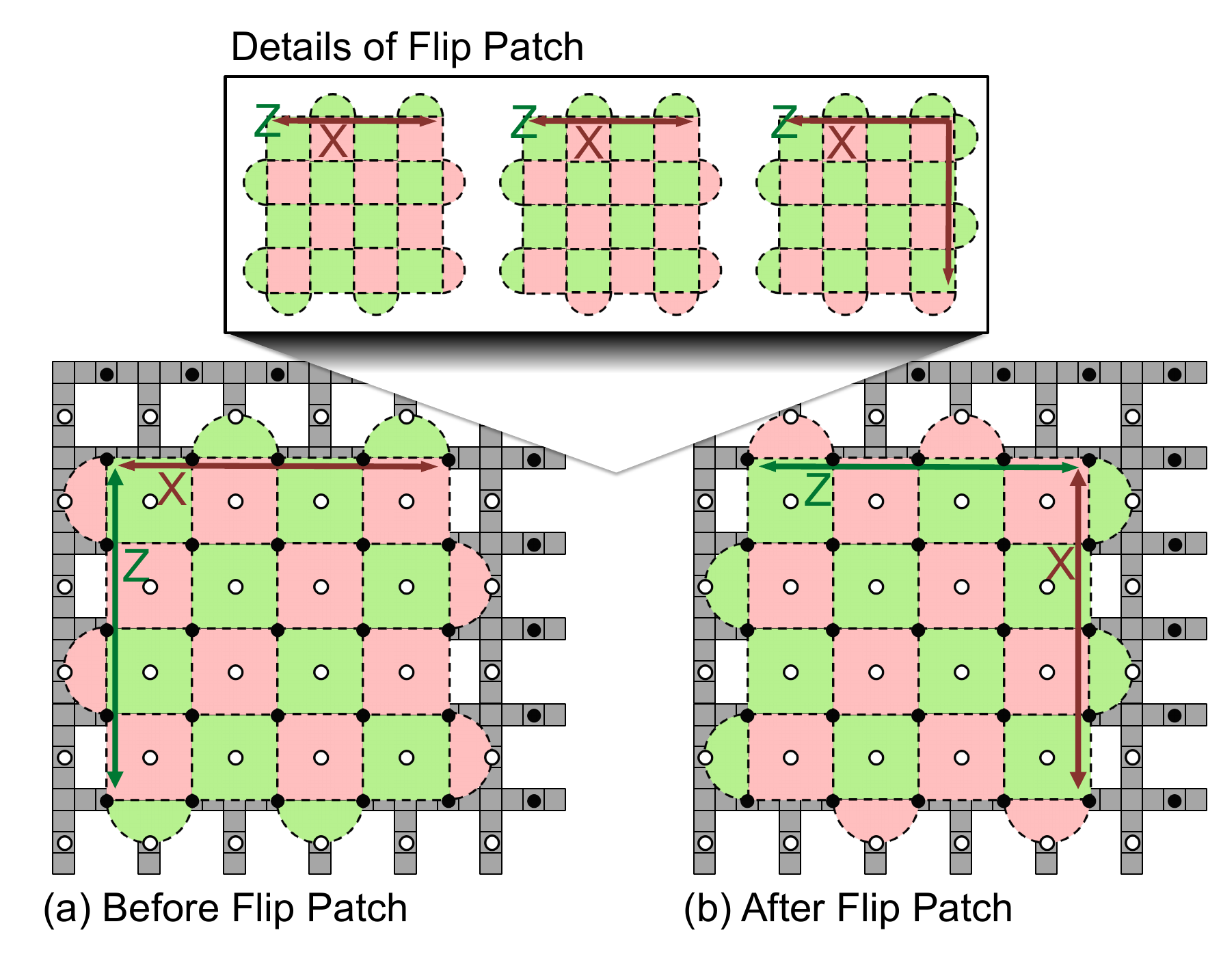}
    \captionsetup{skip=10pt}
    \caption{The Flip Patch operation is a particular sequence of four corner movements that was used as a proxy for the verification of corner movements in general, though it is not considered one of our primitives. The inset shows the intermediate patch states after the first, second, and third corner movements.}
    \label{fig:flip}
\end{figure}

A note of caution is that, while the corner movements implemented within TISCC are general and have in many cases been verified to preserve the encoded logical state (see Sec.~\ref{sec:verification}), not all valid patch deformations can be implemented fault-tolerantly~\cite{vuillot2019code}. For example, consider the Flip Patch operation, an operation implemented within TISCC that can take a patch from the standard arrangement [Fig.~\ref{fig:canonical_arrangements}(a)] to the flipped arrangement [Fig.~\ref{fig:canonical_arrangements}(c)] or from the rotated arrangement [Fig.~\ref{fig:canonical_arrangements}(b)] to the rotated-flipped arrangement [Fig.~\ref{fig:canonical_arrangements}(d)] by performing a sequence of four clockwise corner movements (see Fig.~\ref{fig:flip} for details). This operation was used as a proxy to verify our corner movement functionality and ensure its robustness to different combinations of code distances, but may not be fault-tolerant by the criteria of Ref.~\cite{vuillot2019code}.

\begin{figure}[htbp]
    \centering
    \includegraphics[width=\columnwidth]{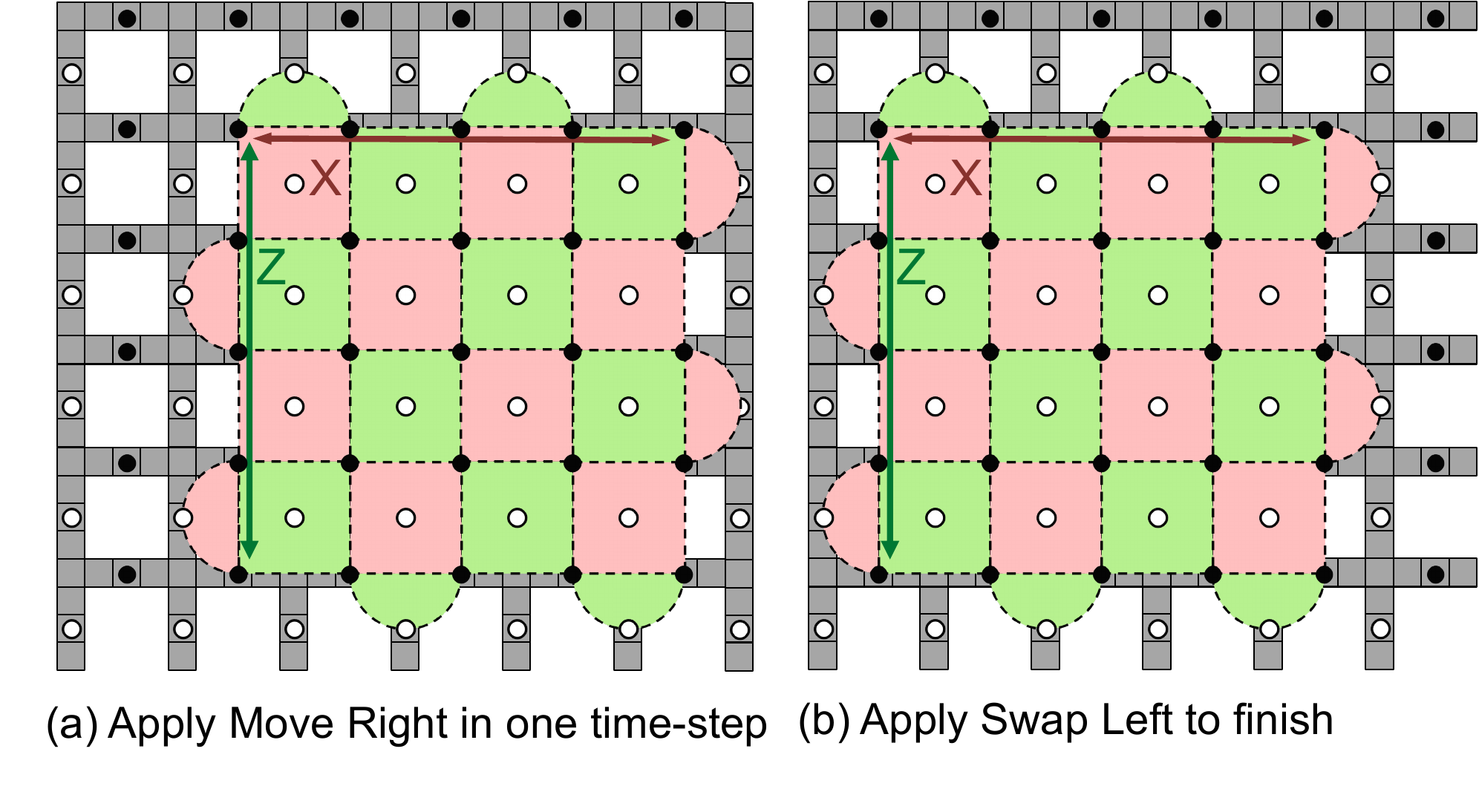}
    \captionsetup{skip=10pt}
    \caption{Move Right (a) and Swap Left (b) are primitives that can be performed in sequence to fault-tolerantly map between the Standard and Rotated-Flipped arrangements (shown) or the Rotated and Flipped arrangements (not shown) in one logical time-step on one logical tile. The Standard arrangement is the implied precursor in (a).}
    \label{fig:move_right_swap_left}
\end{figure}

In addition to patch deformations, since the ability to translate patches without resorting to lattice surgery is often required by authors proposing protocols for patch rotation~\cite{beverland2022surface, fowler2018low}, we have implemented and verified a primitive called Swap Left that relies on trapped-ion movement alone to move data qubits left-ward, effectively `swapping' them with the ancilla strip residing to the right on their respective logical tile. When Swap Left follows a Move Right operation\footnote{Move Right is another verified primitive available in TISCC; it simply performs a one-column move operation to the right, and is considered a primitive in itself since our Move operation (a derived instruction, see Table~\ref{tab:surf_code_derived_instr}) consists of movement over an entire tile.}, a patch can fault-tolerantly be transferred between the standard [Fig.~\ref{fig:canonical_arrangements}(a)] and the rotated-flipped [Fig.~\ref{fig:canonical_arrangements}(d)] or between the rotated [Fig.~\ref{fig:canonical_arrangements}(b)] and the flipped [Fig.~\ref{fig:canonical_arrangements}(c)] arrangements in one logical time-step and on a single tile\footnote{Technically, Move Right requires a tile to borrow a column from the tile to the right of itself to support syndrome measurement qubits for the resultant boundary stabilizers. This is acceptable as long as the adjacent tile, if initialized, does not enter the flipped or rotated-flipped arrangements during this time-step.}. For the sake of brevity, we do not provide detail on the pattern of movements used within Swap Left. See Fig.~\ref{fig:move_right_swap_left} for details of the Move Right and Swap Left operations.

To conclude Sec.~\ref{sec:surface_codes}, we remind the reader that TISCC focuses on what we have called the second step of surface code compilation, in which surface code instructions are compiled into hardware instructions. We have reviewed the literature on the first step of surface code compilation and determined that a local, tile-based lattice surgery instruction set suits our purposes of using verified primitive surface code operations to compile and estimate resources for a limited yet universal set of surface code operations. For a compiler that focuses on the first compilation step and can be configured to use the same (local, tile-based) surface code instruction set in the fault-tolerant layer as TISCC, see the Lattice Surgery Compiler~\cite{watkins2023high}\footnote{In particular, see the May 2023 release of the Lattice Surgery Compiler, which was co-written with TISCC as part of a broader end-to-end resource estimation effort for fault-tolerant implementations of quantum algorithms.}. Next, we move into our discussion on how the surface code is mapped onto a trapped-ion hardware model.

\section{Surface Code Mapping Onto Trapped-Ion Hardware Model}
\label{sec:hardware_model}
\subsection{Trapped-Ion Hardware Architecture}
\label{sec:hardware_grid}
We have loosely based our trapped-ion hardware model off of Quantinuum's system model H1~\cite{quantinuumH1}, which uses the quantum charge-coupled device (QCCD) architecture, and the Sandia National Laboratories Phoenix ion trap~\cite{revelle2020phoenix}. Surface code patches are mapped onto an arbitrarily large rectangular grid of trapping zones, with three trapping zones per straight segment and straight segments connected by junctions. In TISCC, where we consider an abstract representation of a trapped-ion architecture, we will refer to these elements of the grid as quantum sites, or \textit{qsites} for short. See Fig.~\ref{fig:standard_arrangement} for a representation of the grid of `M' (memory), `O' (operation), and `J' (junction) qsites. The architecture as a whole can be thought of as grid-like, where a repeating unit $\{`M', `O', `M', `J', `M', `O', `M'\}$ (two straight segments, one pointed down-ward and one pointed right-ward, connected by a junction) is tiled over the plane.

\subsection{Trapped-Ion Instruction Set and Hardware Parameters}
\label{sec:trapped_ion_instr_set}

\begin{figure}[htbp]
    \begin{multicols}{2} 

    \begin{minipage}{\linewidth}
        \centering
        \captionsetup{skip=10pt}
        \caption{Native trapped-ion gate set implemented within TISCC.}
        \begin{tabular}{|l|c|}
            \hline
            \textbf{Operation} & \textbf{Time ($\mu$s)} \\
            \hline
            Prepare\_Z & 10 \\
            Measure\_Z & 120 \\
            X\_pi/2 & 10 \\
            X\_pi/4 & 10 \\
            Y\_pi/2 & 10 \\
            Y\_pi/4 & 10 \\
            Z\_pi/2 & 3 \\
            Z\_pi/4 & 3 \\
            Z\_pi/8 & 3 \\
            ZZ & 2000 \\
            Move & 5.25 \\
            Junction & 105 \\
            \hline
        \end{tabular}
        \label{tab:tiscc_gate_set}
    \end{minipage}
    
    \begin{minipage}{\linewidth} 
        \centering
        \includegraphics[width=\linewidth]{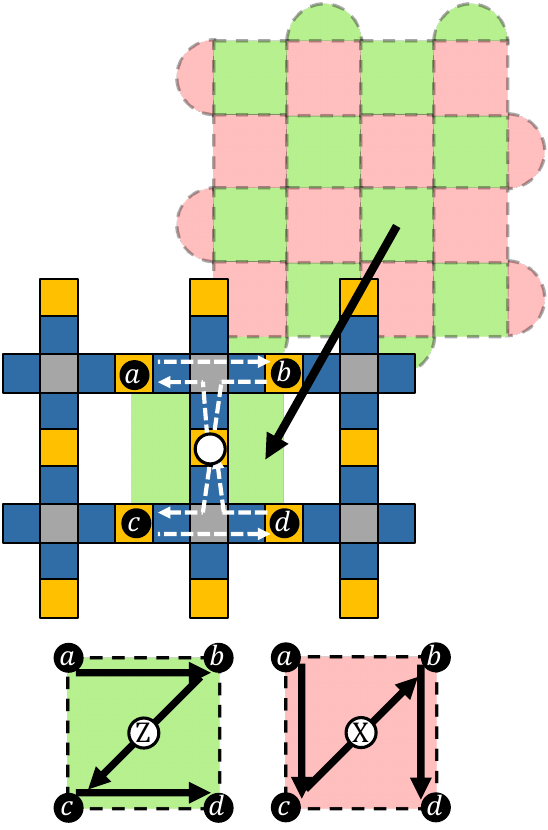}
        \captionsetup{skip=10pt}
        \caption{To interact, the measure qubit (white) moves adjacent to each data qubit (black) in a sequence specified by the (Z or N) measurement pattern.}
        \label{fig:circuit_mapping}
    \end{minipage}
    
    \end{multicols}
\end{figure}

In QCCD trapped-ion compilation, where qubits are mobile, it is sensible to consider trapping zones to be the target acted upon by hardware instructions instead of physical qubits.  The native trapped-ion gate set that is implemented within TISCC is given in Table~\ref{tab:tiscc_gate_set}. Because TISCC is specialized to surface code compilation, which primarily involves circuits composed of the Clifford+T gate set and not arbitrary rotations, we have chosen to specialize to $P_{\theta} = e^{-iP\theta}$, where $P \in \{X, Y, Z\}$ are the Pauli operators and $\theta \in \{\pi/2, \pm \pi/4, \pm \pi/8\}$ are particularly useful angles. For comparison, the H1 documentation~\cite{quantinuumH1} defines a native arbitrary single-qubit rotation gate in the X-Y plane as $U_{1q}(\theta, \phi) = e^{-i(\cos \phi X + \sin \phi Y)\theta/2}$ and a native arbitrary Z-axis rotation as $R_z(\lambda) = e^{-i Z \lambda/2}$. It is worth noting that our $X_{\theta} = U_{1q}(2\theta, 0)$, $Y_{\theta} = U_{1q}(2\theta, \pi/2)$, and $Z_{\theta} = R_z(2\theta)$. The ZZ operation is actually $(ZZ)_{\pi/4}$. We build Hadamard and CNOT gates out of these native gates according to Ref.~\cite{quantinuumH1}. 
 
TISCC explicitly handles movement operations that cause qubits to travel between adjacent trapping zones or through junctions. The Move and Junction operations in Table~\ref{tab:tiscc_gate_set} correspond with per-site movement between trapping zones and to/from junctions, respectively. The timings of the Move and Junction operations are calculated using velocities of 80 m/s~\cite{clark2023characterization} and 4 m/s~\cite{burton2022transport}, respectively, and a trapping zone width of 420 $\mu$m. Since we enforce the constraint that ions cannot sit at junctions, a Junction move will appear in compiled output as Move qsite1 qsite2, where qsite1 and qsite2 are two trapping zones adjacent to the same junction, and time is allocated for 2 Junction move operations. 

In the current verison of TISCC, the Split, Merge, and Cool operations that are typically required during the ZZ operation are implied to take place during the time allocated to the ZZ operation. Here, we assume that the Split, Merge, and Cool operations ($\approx2$~ms) dominate the ZZ operation time ($\approx25$~$\mu$s) with times generalized from Ref.~\cite{pino2021demonstration}. It is anticipated that these operations, as well as qubit loss, will be handled explicitly in future versions to improve the realism of TISCC.

\subsection{Explicit X and Z Stabilizer Circuit Mappings}
\label{sec:circuit_mappings}
Each plaquette is serviced by one mobile syndrome measurement qubit that moves to interact with data qubits in a specified sequence given by the standard surface code syndrome extraction circuits~\cite{fowler2012surface}; see Fig.~\ref{fig:circuit_mapping} for details of this sequence. For a given round of error correction over a patch, the syndrome extraction circuits are performed in parallel over all its plaquettes. In order to prevent errors on syndrome measurement qubits from becoming two data qubit errors parallel to the logical operator of the same kind, thus reducing the effective code distance, we employ two different movement patterns for the ``measure'' qubit: Z pattern and N pattern. This problem (and solution) has been noted elsewhere~\cite{tomita2014low, litinski2018lattice, fowler2018low, chamberland2022circuit}. In general within TISCC, Z-type stabilizers are measured using the Z pattern and X-type stabilizers are measured using the N pattern. Exceptions to this rule occur when the patch is set to either the rotated or flipped arrangement [see Fig.~\ref{fig:canonical_arrangements}(b) and (c)] following a transversal Hadamard operation or the Flip Patch operation, respectively, since in those cases the logical operators change direction. If Flip Patch is followed by the transversal Hadamard [leaving the rotated-flipped arrangement, Fig.~\ref{fig:canonical_arrangements}(d)], the circuit patterns are reset to the standard ones. Other deviations from this rule could occur following corner movements as described in Ref.~\cite{fowler2018low}. This level of stabilizer circuit generality has not yet been implemented in TISCC.

There are occasions where the parallel operation of adjacent stabilizer patterns leads to junction conflicts. In TISCC, we implement basic hardware validity checks such as that two qubits do not move through the same junction at the same time, and that two qubits do not occupy the same site at the same time\footnote{Though trapped-ion systems typically can have qubits occupying the same trapping zone, we do not make us of this in our hardware model.}. When the TISCC hardware validity checker detects a junction conflict, it resolves it by inserting appropriate time to perform the conflicting junction moves sequentially.
 
\label{sec:twists}
Explicit hardware circuits for extended stabilizers and twist defects have not been devised because they are not required by our primitives. The most general lattice surgery instruction set, which includes the measurement of arbitrary Pauli products over the support of the logical qubit register, requires these~\cite{litinski2018lattice, litinski2019game}. The implementation of these defects in lattice surgery protocols is not well-understood due to challenges related to syndrome extraction circuit scheduling and decoding algorithms~\cite{chamberland2022universal, chamberland2022circuit}. Avoiding these complexities was one of the motives for our instruction set choice. However, since there is an ongoing debate about whether a twist-based instruction set would improve or worsen the overhead of fault-tolerance~\cite{chamberland2022universal,chamberland2022circuit}, and given our considerations about extensibility being necessary for a good compiler from Sec.~\ref{sec:introduction}, we believe that it may be pertinent to implement twist-based protocols in future versions of TISCC. 

\subsection{Hardware Resource Estimation}
\label{sec:resource_estimation}

In addition to being a compiler, TISCC is a resource estimator. The circuits output by TISCC are time-resolved, incorporating the values from Table~\ref{tab:tiscc_gate_set} and considering operations that are done in parallel. Thus, execution time is explicitly calculated for each patch operation. Using the master hardware circuit for a given operation, resources such as grid area (in $m^2$), computation time (in $s$), space-time volume ($s*m^2$), number of trapping zones, trapping zone-seconds, and active trapping zone-seconds are calculated.

\section{Verification Of Logical State Using ORQCS}
\label{sec:verification}

In order to verify that hardware circuits output by TISCC correctly correspond with their intended transformations of the logical sub-space, we have employed the Oak Ridge Quasi-Clifford Simulator (ORQCS) to perform (near-)Clifford simulation of these circuits and collect expectation values of both Pauli strings (corresponding with logical operators) and simulated measurement outcomes. ORQCS implements a parser and hardware model for the TISCC instruction set so that the TISCC circuits, written in terms of gates acting on qsites residing on the trapped-ion hardware, are interpreted as unitary operations acting on a quantum state. These results are used together in quantum state and process tomography protocols to verify that final density matrices (in the case of state preparation circuits) and process matrices (in the case of operations that can act on arbitrary encoded states) agree with expectations\footnote{All verification is performed in the absence of simulated hardware errors.}. This procedure has been used to verify our primitives (including those from Table~\ref{tab:surf_code_prim} and others discussed in Sec.~\ref{sec:rotation}), some combinations of primitives representing members of our instruction set from Table~\ref{tab:surf_code_instr_set}, and members of the derived instruction set that is discussed in Sec.~\ref{sec:derived}. The verification of some operations required us to incorporate simulated measurement outcomes in classical post-processing to recover correct values of logical operators. This is often relevant when measurements on data qubits are made during circuit execution (e.g. in Merge, Split, and Measure X/Z) and when corners are moved and boundary stabilizers are modified (e.g. in Flip Patch). We follow Ref.~\cite{nielsen_chuang_2010} for the tomography protocols. Verification has been performed for various even and odd code distances $\geq 5$, including cases where $d_x = d_z$ and cases where $d_x \neq d_z$.

\subsection{Clifford (and Near-Clifford) Simulation}
Verification of non-Clifford circuits has additional overhead. ORQCS computes expectation values of observables by taking a weighted average over a number of samples, where each sample is itself a Clifford circuit. Each non-Clifford gate is represented by a decomposition of Clifford gates, and in each sample, only one of these Clifford gates is randomly chosen to be simulated. The probability that a Clifford gate is selected is determined by the decomposition coefficients, and the weight of the sample is adjusted based on the probability of the selected Clifford gate. Thus the expectation value is computed via a Monte Carlo process, and a number of samples (depending on the number of non-Clifford gates) is required to converge. For our purposes, the only non-Clifford circuit that we verify is a T-state injection circuit that contains a single non-Clifford gate. In this case, verification is done statistically, not exactly, due to the Monte Carlo nature of our simulation technique.

\subsection{Verification of State Preparation Circuits}
\label{sec:state_verif}
We have verified our state preparation primitive circuits (Prepare Z and Inject Y/T) using quantum state tomography within the logical subspace of a single surface code patch; i.e. we construct single-qubit density matrices corresponding with the final logical state using the simulated expectation values of logical operators according to the procedure in Ref.~\cite{nielsen_chuang_2010}. We obtain the correct results for these circuits both with and without the subsequent round of syndrome extraction needed to produce a quiescent state of the surface code~\cite{fowler2012surface} using all four canonical arrangements from Fig.~\ref{fig:canonical_arrangements}. In the absence of hardware errors, the final round of syndrome extraction does not change the result since encoded logical states are unaltered by syndrome extraction~\cite{fowler2012surface}. We note that the Prepare X operation from Table~\ref{tab:surf_code_instr_set} does not have a corresponding primitive because it is composed of Prepare Z followed by a Hadamard. Further verification of the Idle operation, which is composed of $d_t$ rounds of syndrome extraction, is given in Sec.~\ref{sec:one_tile_verif}. To verify the Bell State Preparation instruction (Table~\ref{tab:surf_code_derived_instr}) we used state tomography within the two-qubit logical sub-space and made classical corrections dependent on measurements performed during the merge and split. 


\subsection{Verification of One-Tile Operations}
\label{sec:one_tile_verif}
To verify the effect of one-tile operations on the encoded logical sub-space, we used a single-qubit quantum process tomography protocol and used our verified state preparation circuits to input encoded logical states. All but one of our one-tile operations were verified starting from all of the canonical stabilizer arrangements. As mentioned in Sec.~\ref{sec:rotation}, Flip Patch, though not one of our primitives, was used as a proxy to verify our corner movement functionality. When applied to even, odd, and mixtures of even and odd code distances, Flip Patch covers all of the situations that can arise during corner movements, especially the removal and re-preparation of corner qubits. Flip Patch also serves as an exemplar for the re-construction of logical operators in post-processing since neither of the default logical operators have any overlap in support with their previous selves. In contrast to other operations, Flip Patch was only verified from the standard and rotated arrangements since those are the only arrangements from which it was implemented. See Fig.~\ref{fig:flip} for details of its implementation.

Several of our primitives were expected (and verified) to yield a process matrix that is consistent with the identity process. This is the case for Idle, Flip Patch, Swap Left, and Move Right. Since the Idle operation, which is actually a round of syndrome extraction, lies at the heart of TISCC, we additionally verified for patches as large as $d=30$ that measurement outcomes were stable upon repeated applications of it, which is consistent with the expectation that the surface code is in a quiescent state after the first round of syndrome extraction~\cite{fowler2012surface}. For a more low-level verification of the Idle operation, we verified by hand, for $d=2$, that the stabilizer generators are identical to our expectations after each layer of the parallelized syndrome extraction circuits (including three parallel syndrome extraction circuits in this case) was executed. Specifically, ORQCS outputs a set of generators for the stabilizer group after the application of each layer (e.g. after a couple of parallel CNOT gates are performed). We then generate the whole stabilizer group from these, and check by hand that the expected stabilizer generators are present in that group. 
To verify the Hadamard and Pauli X/Y/Z primitives, we ensured that the process matrices (which were simulated exactly) had the expected non-zero entries.


\subsection{Verification of Two-Tile Operations}

The instruction set in Table~\ref{tab:surf_code_instr_set} contains only two members that operate on two tiles, namely Measure XX/ZZ. For both of these instructions, quantum process tomography on the two-qubit logical sub-space was used for verification, and verification was performed statistically because each instruction has two branches depending on measurement outcomes during the Merge operation (see Ref.~\cite{de2020zx}). In order to individually verify the primitives that compose this operation (Merge and Split), we chose not to explicitly construct process matrices that map two logical qubits to one logical qubit (in the case of Merge) or one logical qubit to two logical qubits (in the case of Split). Instead, we verified by inspection that the resultant density matrices were as expected for an informationally-complete set of input states using the conditional mapping of the logical sub-spaces set forth in Ref.~\cite{de2020zx}. These primitives were also verified within the context of derived instructions; namely, both the patch contraction and patch extension sub-instructions from Table~\ref{tab:surf_code_derived_instr} were verified to be identity processes using single-qubit process tomography. We note that, because Merge and Split were only implemented for patches in the standard stabilizer arrangement, this is the only case for which they were verified. 

\subsection{Post-Processing with Measurements}
\label{sec:post-processing}
While certain stabilizer measurement outcomes are initially non-deterministic, in the absence of noise they are deterministic upon repeated cycles. Also, while the expected values of Pauli strings representing logical operators on the surface code depend on this initial set of measurement outcomes, one typically tracks the \textit{Pauli frame}~\cite{riesebos2017pauli} to reconstruct logical operators \textit{post hoc}. Ref.~\cite{fowler2018low} describes how to use the Pauli frame to adapt the expected values of logical operators by products of stabilizers using a process called \textit{operator movement}. Methods have been incorporated within TISCC to aid the user in tracking operator deformations and operator movements for use in post-processing logical operators using measurement outcomes. 

We can look to the Flip Patch and Move Right primitives for two concrete examples. To aid in the former case, upon modifying boundary stabilizers, TISCC stores the qsite indices pertaining to logical operator deformations within the member variables of the LogicalQubit object, from which they can later be extracted. To aid in the latter case, TISCC contains operator movement functionality wherein one can specify a logical operator and a number of rows or columns to shift and it returns all of the qsites corresponding with the stabilizer measurements needed to deform the operator. Both of these examples show how TISCC gives users the needed information to combine measurement outcomes with expectation values of logical operators to obtain correct results. Similar methods to these are used to verify the the Patch Contraction sub-instructions. Thus, TISCC and ORQCS together provide a valuable platform for simulating and verifying surface code operations.

\section{Conclusion} \label{sec:conclusion}

In conclusion, we have written a software library that produces trapped-ion hardware circuits for a universal lattice surgery instruction set. Since it keeps track of parallel operations and the nominal time at which each hardware gate is applied, it is additionally useful as a resource estimator for surface code operations on trapped-ion processors. Because TISCC constructs a surface code instruction set out of verified primitives, it is known that the circuits it produces are correct. In addition to the instruction set presented here, the verified primitives available within TISCC could aid in the compilation of alternative instruction sets. While TISCC has been specialized to surface codes and trapped-ion processors, we note that it was designed to be modular enough to be extensible to other grid-like hardware architectures and QECC implementable on those architectures. 

In the near future, we foresee that TISCC circuits could be used to produce error rates for each member of our instruction set and could therefore lend themselves to logical error rate estimates for entire circuits when combined with output from lattice surgery compilation tools such as the Lattice Surgery Compiler~\cite{watkins2023high}. While TISCC is currently a proof-of-concept compiler with a limited output instruction set of verified operations involving one to two logical tiles, TISCC can in principle be used to compile circuits for larger systems of logical qubits. Since the primitives have been verified to be correct, combinations of the primitives applied in serial or parallel to non-overlapping sets of surface code patches should also be correct, with only minor caveats. This goes beyond what we have tested in the preparation of this manuscript. 

In the farther future, there are a few directions in which we can foresee TISCC evolving: (i) building a better trapped-ion surface code compiler through (a) a more realistic trapped-ion instruction set (including explicit split, merge, swap, and cool operations), (b) implementations of surface code operations that take better advantage of the expediencies presented by the QCCD architecture (e.g. using ion shuttling instead of lattice surgery for patch movements and rotations), and (c) broadening the lattice surgery instruction set to include Y-basis measurements by implementing twist defects and extended stabilizers as mentioned in Sec.~\ref{sec:twists}, (ii) extending TISCC to other hardware models such as 1D QCCD architectures for near-term applications (e.g. Quantinuum's H1 and H2 systems) or even to other modalities such as superconducting qubits, and (iii) improving the TISCC API for better usage as a programming language for surface code quantum computing. 

\begin{acks}
This work was completed as part of the Defense Advanced Research Projects Agency Quantum Benchmarking program. This manuscript has been authored by UT-Battelle, LLC, under contract DE-AC05-00OR22725 with the US Department of Energy (DOE). The US government retains and the publisher, by accepting the article for publication, acknowledges that the US government retains a nonexclusive, paid-up, irrevocable, worldwide license to publish or reproduce the published form of this manuscript, or allow others to do so, for US government purposes. DOE will provide public access to these results of federally sponsored research in accordance with the DOE Public Access Plan (https://www.energy.gov/downloads/doe-public-access-plan).
\end{acks}

\begin{table*}[htbp]
  \centering
  \caption{`Derived' local lattice surgery instruction set implemented using TISCC primitives. Italicized instructions are not considered part of our instruction set but are included for reference as sub-circuits used by other members.}
  \begin{tabular}{|l|p{8cm}|c|c|}
    \hline
    \textbf{Operation} & \textbf{Description} & \textbf{Logical Tiles In/Out} & \textbf{Logical Time-Steps} \\
    \hline
    Bell State Preparation & Initializes a Bell state on two adjacent uninitialized tiles & 2 & 1 \\
    Bell Basis Measurement & Performs a destructive Bell basis measurement on two adjacent (initialized) tiles and makes uninitialized & 2 & 1 \\
    Extend-Split & Composed of a patch extension followed by a split operation & 2 & 1 \\
    Merge-Contract & Composed of a merge operation followed by a patch contraction & 2 & 1 \\
    Move & Composed of a patch extension followed by a patch contraction & 2 & 1 \\
    \textit{Patch Contraction} & Contracts one initialized two-tile patch into a one-tile patch while    preserving the encoded state & 2/1 & 0 \\
    \textit{Patch Extension} & Extends one initialized one-tile patch into a two-tile patch while preserving the encoded state & 1/2 & 1 \\
    \hline
  \end{tabular}
  \label{tab:surf_code_derived_instr}
\end{table*}

\bibliographystyle{ACM-Reference-Format}
\bibliography{refs}

\appendix

\section{Derived Instruction Set}
\label{sec:derived}

Besides the minimal instruction set (Table~\ref{tab:surf_code_instr_set}), TISCC implements several other instructions (Table~\ref{tab:surf_code_derived_instr}). While these additional instructions could be implemented using instructions from Table~\ref{tab:surf_code_instr_set}, TISCC implements them more efficiently in terms of primitive operations (Table~\ref{tab:surf_code_prim}) by exploiting commutation of stabilizers. For example, a Prepare X operation on one tile followed by a Measure ZZ operation between that tile and an adjacent one need not be performed in two time steps because the logical $\ket{+}$ state need not be fault-tolerantly encoded before the Measure ZZ operation is performed. Because of this, we lump these two operations together into a two-tile operation called Extend-Split. That these types of expediencies are possible is well known~\cite{litinski2019game, beverland2022surface}. See Table~\ref{tab:surf_code_derived_instr} for the list of these instructions and their description as implemented by TISCC.

\section{Software Design and Implementation}\label{sec:software_design}

\subsection{Class Structure}

TISCC can either be compiled into an executable and given command line input (code distances, operation of interest) or used as a library. Important TISCC classes include:

\begin{itemize}
    \item \textbf{GridManager}: Provides access to an array representation of the trapped-ion architecture along with functions to help navigate it. Enforces validity of the final hardware circuit by tracking qubit movement.

    \item \textbf{Plaquette}: Primarily tracks the grid indices (qsites) occupied by the qubits supported by a stabilizer plaquette.

    \item \textbf{LogicalQubit}: Constructed by requesting Plaquettes from the \texttt{GridManager}. Provides functions to compile the patch-level operations mentioned in Sec.~\ref{sec:introduction}. Manages its Plaquettes, parity check matrix, and logical operators by updating them when necessary and testing validity.
    
    \item \textbf{HardwareModel}: Defines a set of native hardware operations and related parameters. Compiles gates requested by \texttt{LogicalQubit} to the native gate set and adds native gates to a time-resolved hardware circuit.
\end{itemize}

To use TISCC, one typically initializes the GridManager with the size of the hardware grid.  Then, LogicalQubit(s) are added.  Finally, primitive operations from Table~\ref{tab:surf_code_prim} are appended using the appropriate LogicalQubit methods. Lastly, validity of the hardware circuit is enforced through the GridManager and the circuit and/or final resource counts are printed.

\end{document}